%% file: main.tex
\documentclass[11pt]{article}

\usepackage{structure}
\usepackage{subcaption}
\usepackage{graphicx}

\begin{document}

\input{Sections/Title}
\linespread{1.5}
\input{Sections/Abstract}
\input{Sections/Introduction}
\input{Sections/new_methodology}

\input{Sections/new_Results_and_Discussion}

\input{Sections/Conclusions}
\input{Sections/Acknowledgements}
\input{Sections/Author_Contributions}
\input{Sections/Data_Availability}
\input{Sections/Code_Availability}
\input{Sections/Competing_Interests}
\newpage
\printbibliography[title={References}]

\end{document}

%% file: Sections/Title.tex
\vspace*{2pc}

{\LARGE  \textbf{Advancing state estimation for lithium-ion batteries with hysteresis: systematic extended Kalman filter tuning}}

\vspace{1pc}

\hspace{1pc}{\large J Knox$^{1,2,*}$, M Blyth$^{1,2}$, A Hales$^{1,2}$, }

{\small\hspace{1pc}$^1$Faculty of Engineering, University of Bristol, BS8 1TR, United Kingdom}

{\small\hspace{1pc}$^2$The Faraday Institution, Quad One, Becquerel Avenue, Harwell Campus, Didcot OX11 0RA, United Kingdom}

{\small\hspace{1pc}$^*$Corresponding author}

\vspace{3pc}

\hspace{1pc}{J Knox ORCiD: 0009-0009-9795-4989}

\hspace{1pc}{M Blyth ORCiD: 0000-0001-8890-7800}

\hspace{1pc}{A Hales ORCiD: 0000-0001-6126-6986}

\vspace{3pc}

%% file: Sections/Abstract.tex
\section*{Abstract}

Knowledge of remaining battery charge is fundamental to electric vehicle deployment.
Accurate measurements of state-of-charge (SOC) cannot be directly obtained, and estimation methods must be used instead.
This requires both a good model of a battery and a well-designed state estimator.
Here, hysteretic reduced-order battery models and adaptive extended Kalman filter estimators are shown to be highly effective, accurate SOC estimators.
A battery model parameterisation framework is proposed, which enhances standardised methods to capture hysteresis effects.
The hysteretic model is parameterised for three independent NMC811 lithium-ion cells and is shown to reduce voltage RMS error by 50\% across 18-hour automotive drive-cycles.
Parameterised models are used alongside an extended Kalman filter, which demonstrates the value of adaptive filter parameterisation schemes.
When used alongside an extended Kalman filter, adaptive covariance matrices yield highly accurate SOC estimates, reducing SOC estimation error by 85\%, compared to the industry standard battery model.

%% file: Sections/Introduction.tex
\section*{Introduction}
\label{sec:Introduction}

Lithium-ion batteries (\abr{LIB}) are currently the preferred power technology for hybrid and electric vehicles (\abr{EV}) due to their high energy density, efficiency, and extended lifespan \cite{Saw2016}. Advanced battery management systems (\abr{BMS}) are required to enhance the safety and longevity of \abr{LIB} cells, and apply corrective measures when operational limits are exceeded. Accurate real-time estimation of State of Charge (\abr{SOC}) is a critical task for the \abr{BMS}. \abr{EV}s, the dominant application for BMSs in 2023, require robust and accurate \abr{SOC} estimators to maximise battery lifetime performance and range, and to meet consumer demand for reliable driving range predictions~\cite{hannan2017review}.

Direct SOC measurement techniques, such as Coulomb counting, are adversely affected by poor initialisation, drifts resulting from current sensor noise, and variations in cell capacity due to changes in State of Health (\abr{SOH}) and temperature \cite{Nejad}. Open-circuit voltage (\abr{OCV}) may be used to infer \abr{SOC} via lookup tables \cite{Zhang2018_1}. However, long relaxation periods are required to approximate \abr{OCV}, rendering it impractical for online state estimation. The presence of hysteresis also significantly reduces the accuracy of this approach~\cite{zhang2018state}. Data-driven methods such as artificial neural networks \cite{Chen2018} model cell behaviour from measured data, without prior knowledge of cell properties; these machine learning algorithms typically suffer from poor generalisation, and thus require extensive datasets to achieve sufficient accuracy.

Cell SOC may be implicitly derived from accessible measurements such as terminal voltage, current, and surface temperature, using mathematical models. In principle, a selection of models may be deployed for this purpose~\cite{Hu2}. Electrochemical models consider fundamental interactions at the particle scale within the electrodes \cite{Hu2018}. Although capturing the underlying physics, these models require more computational resources than are typically available in a \abr{BMS}, and extensive parameterisation is necessary to operate the models effectively. Empirical models describe the evolution of cell voltage using relationships to measurable parameters, but can suffer from inaccuracies of 5-20\% as key dynamics may not be fully captured~\cite{Nejad}. 

Equivalent circuit models (\abr{ECM}) simulate the response of a cell using basic electrical components, such as voltage sources, capacitors, and resistors. The most commonly employed ECM in BMSs is the Th\'evenin model, typically containing two resistor-capacitor parallel pairs in series with a resistor, and a voltage source representing OCV. The architecture of ECMs is comparatively less complex than physics-based models, making them ideally suited for real-time operation and applications within \abr{BMS}s. Whilst \abr{ECM}s do not describe intrinsic physical processes as with electrochemical models, they are significantly easier to parameterise accurately~\cite{Koirala2015,Naseri2022}. The accuracy of an \abr{ECM} is heavily dependent on the quality of its parameters, so numerous parameterisation methods have been proposed~\cite{Hua2021,Birkl2013}. In all cases, the voltage response of a cell is recorded experimentally, under a particular current-based loading profile. The galvanostatic intermittent titration technique (\abr{GITT}) is typically used to fit ECM voltage responses, with numerical algorithms optimising the ECM component parameters to minimise error between the experimental data and modelled voltage response. Model parameters vary with temperature and SOC, so parameterisation and experiments must explore a suitable a range of temperatures and charges.
Voltage hysteresis is known to cause considerable error in ECMs and subsequent SOC estimation~\cite{barai2015study}, yet the widely adopted Th\'evenin model architecture is unable to account for its effects. Previous studies include additional circuit components to account for hysteresis~\cite{Nejad}, however a standardised time-domain based parameterisation technique for \abr{ECM}s with hysteresis is not reported in the published literature~\cite{Huria,Plett2015}.

In \abr{BMS} applications, cell voltage informs predictions of \abr{SOC} through algorithms such as the extended Kalman filter (\abr{EKF}). Kalman filters combine measured data with a model, to directly compute SOC and other relevant cell state variables. The combination of both modelled and measured data produces a more accurate SOC estimation than could be obtained from either individually. \abr{EKF}s are applied to nonlinear systems, and provide an approximate solution to the optimal state-estimation problem~\cite{Simon2006}.
Appropriate tuning of \abr{EKF} parameters remains a difficult task.
Inadequately tuned filters may produce unstable results, and time-consuming empirical tuning methods, although widely adopted in literature, cannot be generalised to other models~\cite{schneider2013not}.
Systematic determination of EKF parameters for \abr{ECM} applications have been proposed to account for current sensor noise and model parameter errors~\cite{Rzepka}.
The method is capable of automatically selecting filter parameters to reflect the current operating conditions and model parameters of a cell; this is referred to as an adaptive filtering scheme.
Adaptive \abr{EKF}s have not been demonstrated on experimental data for SOC estimation, and simplicity-accuracy trade-offs remain unexplored.

This work sets out a time-domain-based approach to \abr{ECM} parameterisation with hysteresis; the inclusion of a hysteresis element within an ECM is shown to yield a significant improvement in the model's accuracy. Further, a method to systematically tune EKF parameters for hysteretic battery models is experimentally validated. The use of dynamic EKF parameters is shown to increase the accuracy of \abr{SOC} estimation, when validated on experimental measurements. The parameterisation process and associated systematic filter tuning methods presented will allow \abr{BMS} manufacturers to improve the accuracy of \abr{SOC} estimation, and refine tolerances on cell operational limits. This has the potential to increase accessible energy for a battery, and to enable enhanced battery control algorithms to mitigate operational scenarios which cause rapid degradation. Future research directions are proposed, to further enhance hysteretic SOC estimation.

%% file: Sections/new_methodology.tex
\section*{Methods}
\label{sec:Methodology2}

The following section details the hysteresis model architecture, model parameterisation methods and adaptive Kalman Filter tuning methodology.

\subsection*{Modelling}

\begin{figure}
    \centering
    \resizebox{9cm}{!}{\input{Figures/ECM}}
    \caption{Proposed enhanced self-correcting equivalent circuit model. Includes an ideal voltage source $v_{oc}$, equivalent series resistance $R_0$, two resistor-capacitor (\abr{RC}) networks and hysteresis voltage element $v_H$ \cite{Plett2004_2}.}
    \label{fig:ESC}
\end{figure}
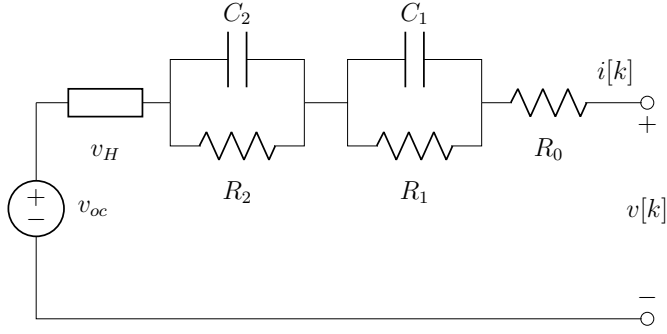

This study benchmarks expected ECM performance using a standard Th\'evenin model with two resistor-capacitor parallel pairs in series with a resistor.
The proposed enhancement to the standard Th\'evenin model, tested for suitability in this study, is the enhanced self-correcting model (\abr{ESC}) introduced by Plett \cite{Plett2004_2}.
This connects the conventional Th\'evenin model in series with a single-state hysteresis voltage component $v_H$. 
The ESC model architecture is presented in Figure \ref{fig:ESC}, where $v_{oc}$ represents cell \abr{OCV}, $R_0$ represents the equivalent series resistance and each resistor-capacitor (\abr{RC})-network captures the effects of diffusive electrochemical behaviours, with time constants $\tau_j = R_jC_j$. 
The hysteresis voltage element is defined by $v_H = Mh$, where $M$ is maximum hysteresis voltage, and $h$ is a unitless hysteresis state bounded between $-1$ and $1$.

A discrete-time representation of the \abr{ESC} is necessary for deployment in lightweight \abr{BMS} controllers \cite{Plett2015}. The discretisation scheme requires model inputs to be uniformly sampled in time, and current $i$ to be constant over the sampling period $\Delta t$. The latter necessitates sampling duration \(\Delta t\) to be small relative to the timescale of battery operation. The discrete-time \abr{ESC} state equation is a map describing the evolution of model variables over time index $k$, including \abr{SOC} $z_k$, \abr{RC}-network currents $i_{R_j,k}$, and the hysteresis state $h_k$. Dynamics are given by
\begin{align}
    \label{eqn:ESC state eqn 1}
    z_{k+1} &= z_k -\frac{\eta_k \Delta t}{Q} i_k,\\
    i_{R_j,k+1} &= \alpha_{RC_j} i_{R_j,k} + (1-\alpha_{RC_j}) i_k,\\
    \label{eqn:ESC state eqn 3}
    h_{k+1} &= \alpha_{H,k} h_k + (A_{H,k} - 1) \mathrm{sgn}(i_k),
\end{align}
where $j$ is the index of the \abr{RC}-network, $\eta$ is cell efficiency, $Q$ is cell capacity, $ \alpha_{RC_j} = \mathrm{exp} (-\Delta t / \tau_j)$ is the \abr{RC}-subcircuit rate factor and $ \alpha_{H_k} = \mathrm{exp} (-| i_k \eta  \gamma \Delta t / Q |)$ is the hysteresis evolution factor with rate constant $\gamma$. The \abr{ESC} output equation gives the terminal voltage of the cell $v_k$ at time $k$ as
\begin{equation}
    \label{eqn:ESC output eqn}
    v_k = v_{oc}(z_k,T_k) + Mh_k - \sum_j R_j i_{R_j,k} - R_0 i_k~,
\end{equation}
where $T$ is temperature.
This work assumes \(\eta=1\).

\subsection*{Experimental Setup}

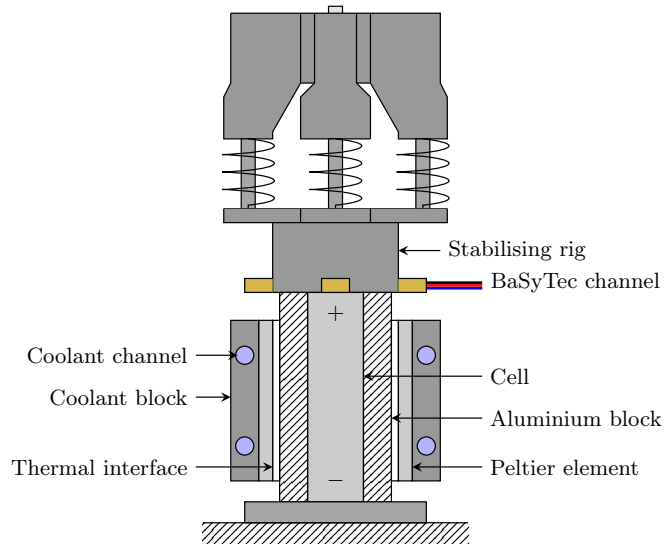
\begin{figure}
    \centering
    \resizebox{9cm}{!}{\input{Figures/CellRig}}
    \caption{Experimental setup for  a single cell connected to a channel on the \textit{BaSyTec CTS} battery cycler (not shown) and thermally controlled using two side-mounted Peltier elements and a thermal controller (not shown)} 
    \label{fig:CellRig}
\end{figure}

All experiments were conducted on 5 Ah \textit{LG Chem M50LT} 21700 cylindrical lithium-ion cells, which comprise an NMC cathode and graphite-silicon anode~\cite{LG2016, ORegan2022} and have previously been used in relevant publications related to \abr{EV} \abr{BMS}s \cite{LeBel,Natella2023}. Three cells were used for the tests, with each cell subjected to all procedures, resulting in three complete datasets. All experiments were conducted using a 4-wire connection \textit{BaSyTec CTS} battery cycler, supporting 1 mV voltage precision, 1 mA current precision~\cite{alvatekCellTest} and a data sampling rate of 10 Hz, which was used throughout. The procedures were conducted at a cell surface temperature of 25 \textdegree C, maintained to within $\pm\;$0.25 \textdegree C using cell-surface mounted Peltier elements driven by thermal control hardware. A schematic of the experimental setup is shown in Figure \ref{fig:CellRig}. Coaxial probes were used to make an electrical connection to the positive terminal and a bespoke annular connector was used to make the negative connection to the `shoulder' of the cell, mimicking battery pack assembly processes used in the EV industry. The stabilisation rig maintained uniform and repeatable pressure applied from the electrical connections onto the cell, ensuring traceability across each test setup and each cell. 

\subsection*{Experimental Procedures}

Coulombic counting is relied upon to extract the real-time SOC from the experimental data.
It was therefore important that all experiments were conducted in isolation, so that any uncertainty in the initial SOC was not able to accumulate over subsequent tests.
To this end, a standardised `SOC reset' procedure was employed between each test, and always ran at 25 \textdegree C.

SOC reset procedure:
\begin{enumerate}
    \item 0.3 C constant current charge to 4.2 V (the upper voltage limit of the cell).
    \item Constant voltage hold at 4.2 V, until $i$<0.01 C.
    \item Rest for 4 hours to ensure electrochemical equilisation is reached.
\end{enumerate}

\begin{figure}

    \begin{subfigure}{0.95\textwidth}
    \centering
    \includegraphics[width=\textwidth]{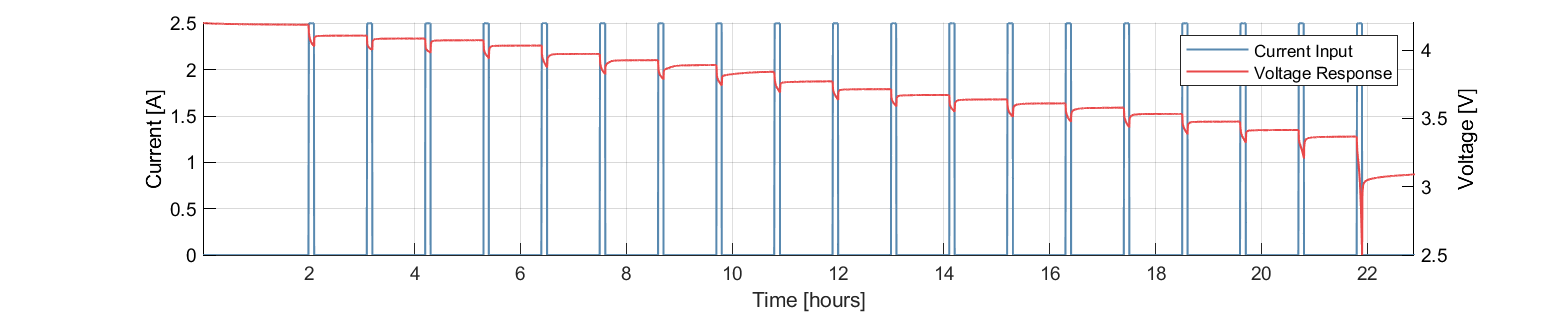}
    \caption{GITT}
    \end{subfigure}

    \begin{subfigure}{0.95\textwidth}
    \includegraphics[width=\textwidth]{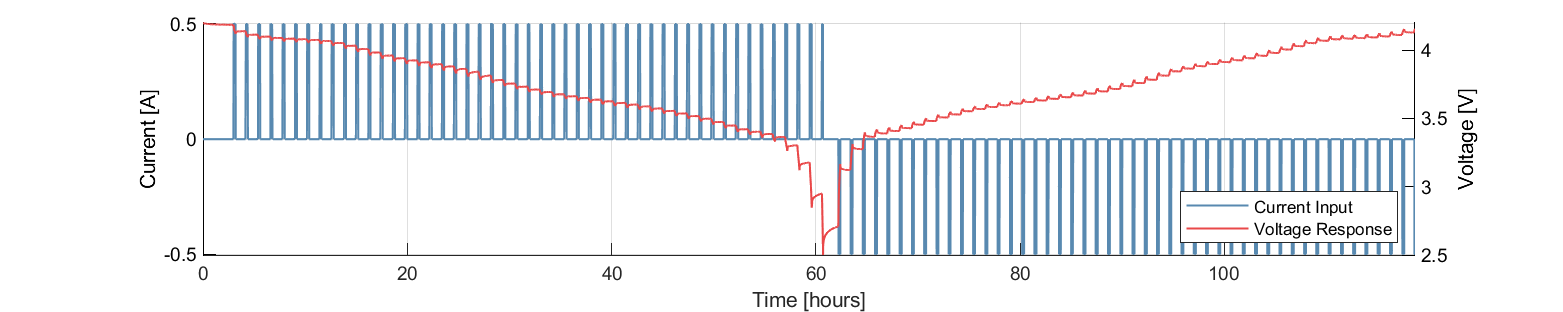}
    \caption{GITT for OCV}
    \end{subfigure}

        \begin{subfigure}{0.95\textwidth}
    \includegraphics[width=\textwidth]{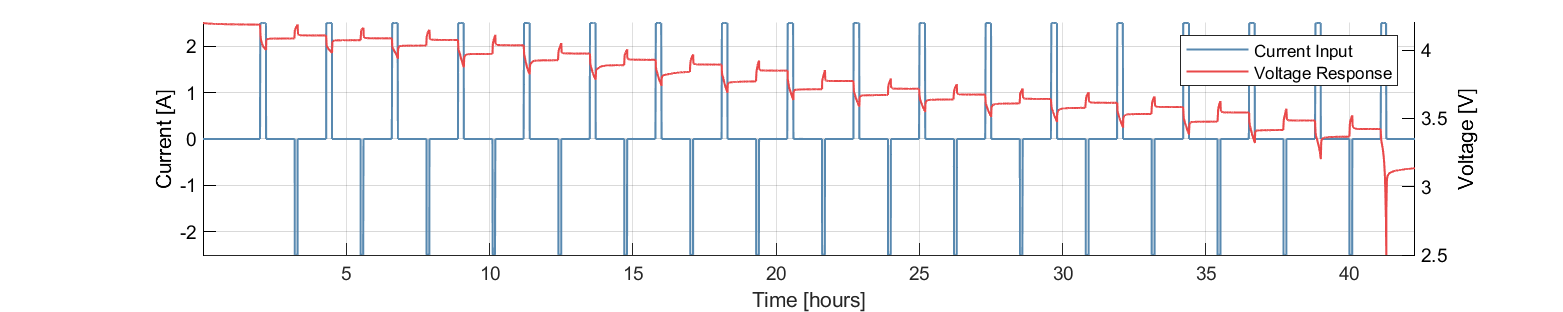}
    \caption{GITT with charge pulse}
    \end{subfigure}

        \begin{subfigure}{0.95\textwidth}
    \includegraphics[width=\textwidth]{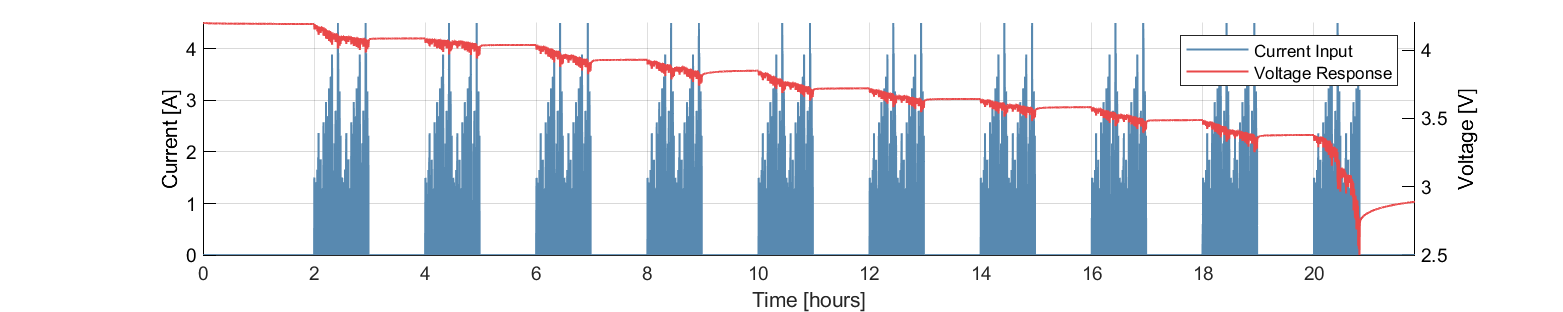}
    \caption{WLTP cycle}
    \end{subfigure}

    \caption{Example experimental data taken directly from each of the four tests completed on each cell. Data from Cell A is shown here. Current ($I$) > 0 A signifies a discharge of the cell.}
    \label{fig:Example experimental data}
\end{figure}

Four experiments were required to parameterise and test the \abr{ESC} model. The following four procedures were conducted on each of the three cells under test:

\begin{itemize}
    \item GITT;
    \item GITT for OCV;
    \item GITT with charge pulse;
    \item WLTP cycle.
\end{itemize}
These procedures are now introduced, with examples from Cell A plotted in Figure \ref{fig:Example experimental data}. 

\textbf{GITT.} A series of constant current discharge pulses and relaxation periods. In this study, 6 minute 0.5 C constant current discharge pulses were used to reduce SOC by 5 \% during each titration, with 1 hour rests between each discharge pulse. The procedure was initiated with the cell at 100 \% SOC, and repeated until the the lower voltage limit of the cell (2.5 V) was reached.

\textbf{GITT for OCV.} Following~\cite{birkl2015parametric}, 12 minute, 0.1 C constant current discharge pulses were used to reduce SOC by 2 \% during each titration, with 1 hour rests between each discharge pulse. The procedure was initiated with the cell at 100~\% SOC, and repeated until the the lower voltage limit of the cell (2.5 V) was reached. The procedure was then reversed, utilising 0.1 C charge pulses, repeated until 4.2 V was reached.

\textbf{GITT with charge pulse.}  An alternating discharge-charge GITT procedure. Discharge consists of a 12 minute 0.5 C constant current discharge pulse to reduce SOC by 10 \%, followed by a 1 hour rest. The cell is then charged for 6 minutes at 0.5 C, increasing SOC by 5 \%. The procedure is initiated with the cell at 100~\% SOC, and repeated until the lower voltage limit (2.5 V) is reached.


\textbf{WLTP cycle.} For model validation and performance quantification, a drive cycle incorporating the Worldwide Harmonised Light Vehicle Test Procedure (WLTP) was used \cite{WLTPFacts}.
To determine the power required for an individual cell from the WLTP, the Driving Cycle block was used within Mathworks' Simulink software \cite{Auger2023Simulink}, alongside simple vehicle dynamics representing an arbitrary EV.
Following the process set out by Hales \textit{et al.} \cite{Hales2021SAE}, 10\% of the battery was discharged over a 1-hour current-based loading profile, followed by a 1-hour rest. 
This procedure was then repeated until the lower voltage limit (2.5 V) was reached. 
This is designed to represent an EV completing a number of 1-hour journeys with short rests in between, allowing the applied validation protocol to evaluate both kinetic and rest-period SOC estimation. 
Whilst a current-based drive-cycle was desirable for the experimental work, the resulting experimental power profile was used for model validation.
This represents real-world conditions whereby Coulomb counting cannot be readily applied.

\subsection*{Parameter Identification}

Cell OCVs are identified from the GITT for OCV experiments. To account for hysteresis, measurements are averaged over the discharge and charge components of the procedure. 




For given RC time-constants, parameters $R_0$, $R_1$, $R_2$, $C_1$ and $C_2$ of the standard Th\'evenin model (without hysteresis) may be identified from GITT data with ordinary least-squares, using established methodologies from the literature~\cite{LeBel,Birkl2015,Nikolian,zhao2018modeling}.
Hysteresis parameters cannot be identified in the same way, and must instead be determined through nonlinear optimisation.
Parameter values of any given ECM component will change when hysteresis is included, due to the overpotential contribution of the hysteresis element, so ESC model parameters cannot be transferred over from a linear least squares parametersiation of an ECM.
Instead, all parameters must be identified through nonlinear optimisation.
Nevertheless, the overpotential contribution of the hysteresis element is small, so ESC (with hysteresis) component parameters will be well-approximated by their ECM (without hysteresis) equivalents.
ECM parameters are therefore be used as an initial solution for the nonlinear parameterisation.
When combined with local gradient-based numerical optimisers, this allows ESC parameters to be identified quickly and accurately.

Initially, Th\'evenin ECM parameters are fitted to GITT data using ordinary least squares.
Time constants \(\tau_1=2\) and \(\tau_2=50\) are chosen based on previous works~\cite{Birkl2013,Samieian2022}.
Parameters of each GITT pulse are used to populate a lookup table (LUT), with linear interpolation used between breakpoints.
The resulting parameters are also used as initial conditions for a nonlinear optimisation procedure. 
A numerical optimisation routine is used to identify parameters $R_0$, $R_1$, $R_2$, $\tau_1 = R_1 C_1$ and $\tau_2 = R_2 C_2$ at each SOC break-point in the LUT, for some given value of \(\gamma\).
The resulting least-squares error quantifies the best-possible ECM model-fit for that value of \(\gamma\).
A further numerical optimisation routine is used, to identify the optimal \(\gamma\) that minimises this error.
GITT with charge-pulse data are used for this, to capture changes in the hysteresis state.

To optimise, hysteresis overpotential \(M\) is first computed from GITT for OCV data as \[M(\mathrm{SOC}) = 0.5\left[v_{oc}^\mathrm{charge}(\mathrm{SOC}) - v_{ov}^\mathrm{discharge}(\mathrm{SOC})\right].\]
Once \(M\) is computed, nonlinear parameter optimisation is performed, using \abr{MATLAB}/Simulink Design Optimisation \cite{MatlabDOTB}.
Simscape electrical components are used to model standard circuit elements.
At each timestep, a Coulomb counter updates \abr{SOC}, and ECM parameter values are extracted from the current optimiser solution-guess.
An output voltage is computed from this, and the associated residual is passed to a numerical optimisation algorithm to optimise component values at any given SOC.
Gradient-based nonlinear least-squares optimisers are preferable, due to fast convergence rates \cite{Jackey2013,Jackey2009}.
Separate estimation tasks are performed at each \abr{LUT} breakpoint.


Parameter bounds are required to ensure a well-posed identification problem~\cite{grandjean2017structural}.
To isolate the effects of each \abr{RC}-network, time constants must be constrained to disjoint ranges.
Here, parameter constraints are selected through trial-and-error, based on the observed transient decay rates in GITT data.
The chosen parameter constraints are
\begin{itemize}
    \item $0 \; \Omega \leq R_0 \leq 1 \; \Omega $
    \item $0 \; \Omega \leq R_1 \leq 1 \; \Omega $
    \item $0 \; \Omega \leq R_2 \leq 1 \; \Omega $
    \item $0.5 \; s \leq \tau_1 \leq 25 \; s $
    \item $50 \; s \leq \tau_2 \leq 500 \; s $
    \item $0 \leq \gamma \leq 3000 $
\end{itemize}
Notably, it was found that if $\tau_1 < 0.5$ s, the optimiser would associate the instantaneously acting `series' resistance with $R_1$ as well as $R_0$, leading to unstable parameters with large variance from one SOC bound to the next.




\subsection*{Kalman filtering for SOC estimation}

Kalman filters and extended Kalman filters have been covered extensively in the literature~\cite{ribeiro2004kalman,terejanu2008extended,schneider2013not,humpherys2012fresh}, including detailed discussions for LIB SOC estimation~\cite{Rzepka,sepasi2014improved,chen2012state,jiang2013extended}.
We follow the EKF implementation proposed in~\cite{Rzepka}; a short summary is provided here.

A state quantifies the instantaneous configuration of a dynamical system.
As per Eqs.~\eqref{eqn:ESC state eqn 1}--\eqref{eqn:ESC state eqn 3}, the state vector of the ESC at time \(k\) is \(x_k = [z_k, i_{R_j, k}, h_k]\).
Given an initial state \(x_0\), a model, and current draw \(i(t)\), the state of the LIB cell at some future time may be computed.
Nevertheless, computed states may diverge from actual system states, due to modelling or sensor errors.
Combining modelled state estimates \(\hat{x}_k^-\) with measured data \(y_k^-\) provides a means to correct this error.
Kalman filters are an optimal state estimation algorithm for achieving this goal.
Filtered state estimates \(\hat{x}_k^+\) provide explicit SOC inferences, and do not require the BMS to perform any post-processing to extract SOC from the results.
Kalman filters require linear system dynamics, and therefore cannot be applied to the ESC, as the hysteresis term is nonlinear in current-draw \(i\).
The extended Kalman filter generalises the method for nonlinear systems, by taking a linear approximation of the system dynamics at each step.
While this does not provide guarantees of optimality, it often works well in practise, when filters are properly tuned~\cite{schneider2013not}.

Uncertainties in the modelled and measured data are quantified through covariance matrices \(\Sigma_{\hat{x},k}^-\) and \(\Sigma_v\), respectively.
Unfiltered data \(\hat{x}_k^-\) and \(\hat{y}_k^-\) are combined and weighted according to their respective uncertainties, to produce a state estimate \(\hat{x}_k^+\) that minimises the \(\ell_2\) state estimation error.
In addition, covariance matrix \(\Sigma_{\hat{x},k}^-\) is also updated in light of available data, giving a covariance matrix \(\Sigma_{\tilde{x},k}^+\) capturing the uncertainty in \(\hat{x}_k^+\).
As noted, SOC is available directly from the state estimate \(\hat{x}_k^+\).

Tuning a Kalman filter involves selecting appropriate values for initial state estimate \(\hat{x}_0^+\), state covariance \(\Sigma_{\tilde{x},0}^+\), measurement noise covariance \(\Sigma_v\), and process noise covariance \(\Sigma_w\).
Accurate uncertainty tuning is crucial for success with extended Kalman filters~\cite{schneider2013not}.
Recent work has demonstrated how filters can be tuned systematically for SOC estimation in hysteretic battery models~\cite{Rzepka}.
Two types of filter are proposed.
An adaptive filter selects measurement and process covariance matrices according to the instantaneous SOC and current-draw, in much the same fashion as an ECM chooses component parameter values for the current SOC; a non-adaptive filter uses the same measurement and process covariance matrices across the entire depth of discharge.
We consider both non-adaptive and adaptive Kalman filters here, implemented using the methodology presented in~\cite{Rzepka}.

%% file: Figures/ECM.tex
\begin{circuitikz}[/tikz/circuitikz/bipoles/length=1.5cm,scale=0.7]

\large

\draw (0,0)  -- (0.7,0)
to[generic,l_=$v_H$] (2.5,0) -- (3.1,0);

\draw (0,0) -- (0,-1.8) to [V,v=$v_{oc}$] (0,-3.1) -- (0,-5) -- (14,-5);

\draw (3.1,0) -- (3.1,1) -- (4.1,1)
to [C, l^=$C_2$] (5.2,1) -- (6.2,1) -- (6.2,0);

\draw (3.1,0) -- (3.1,-1) -- (3.8,-1)
to [R, l_=$R_2$] (5.5,-1) -- (6.2,-1) -- (6.2,0);

\draw (6.2,0) -- (7.2,0) -- (7.2,1) -- (8,1) to [C, l^=$C_1$] (9.5,1) -- (10.3,1) -- (10.3,0);

\draw (7.2,0) -- (7.2,-1) -- (7.9,-1) to [R, l_=$R_1$] (9.6,-1) -- (10.3,-1) -- (10.3,0);

\draw (10.3,0) -- (11,0) to [R, l_=$R_0$] (12.7,0) --(14,0);

\draw (14.13,0) circle (0.15);
\draw (14.13,-5) circle (0.15);
\node at (13.4,0.7) {$i[k]$};
\node at (14.13,-2.5) {$v[k]$};
\node at (14.13,-0.5) {$+$};
\node at (14.13,-4.5) {$-$};
\node at (7,-5.5) {};   
\node at (7,1.5) {};   

\end{circuitikz}

%% file: Figures/CellRig.tex
\begin{tikzpicture}[scale=0.8, every node/.style={transform shape}]
{\footnotesize

\fill[gray!80] (0,0) rectangle (2,-1);
\fill[gray!80] (2,-1.8) rectangle (1.5,-1.2);

\fill[gray!80] (2,-1.2) -- (1.8,-1) -- (1.5,-1) -- (1.5,-1.5) -- (2,-1.2);

\draw[fill=gray!20] (1.4,0) -- (1.4,0.1) -- (1.6,0.1) -- (1.6,0);
\draw[fill=gray!80] (0,0) -- (3,0) -- (3,-1) -- (3.1,-1.2) -- (3.1,-1.8) -- (2.4,-1.8) -- (2.4,-1.7) -- (2,-1) -- (1.8,-1);
\draw (2,-1) -- (2,0);
\draw[fill=gray!80] (0,0) -- (0,-1) -- (-0.1,-1.2) -- (-0.1,-1.8) -- (0.6,-1.8) -- (0.6,-1.7) -- (1,-1) -- (1.2,-1);
\draw (1,-1) -- (1,0);
\draw[fill=gray!80] (1.2,0) -- (1.2,-1) -- (1,-1.2) -- (1,-1.8) -- (2,-1.8);
\draw (1.8,0) -- (1.8,-1) -- (2,-1.2) -- (2,-1.8);

\draw[fill=gray!70] (0.15,-1.8) rectangle (0.35,-2.8);
\draw[decoration={aspect=0.1, segment length=2mm, amplitude=3mm,coil},decorate] (0.25,-1.8) -- (0.25,-2.8);

\draw[fill=gray!70] (2.65,-1.8) rectangle (2.85,-2.8);
\draw[decoration={aspect=0.1, segment length=2mm, amplitude=3mm,coil},decorate] (2.75,-1.8) -- (2.75,-2.8); 

\draw[fill=gray!70] (1.4,-1.8) rectangle (1.6,-2.8);
\draw[decoration={aspect=0.1, segment length=2mm, amplitude=3mm,coil},decorate] (1.5,-1.8) -- (1.5,-2.8);

\fill[gray!80] (0.6,-4) rectangle (2.4,-3);
\draw (-0.1,-2.8) -- (3.1,-2.8);
\draw[fill=gray!80] (-0.1,-2.8) -- (-0.1,-3) -- (3.1,-3) -- (3.1,-2.8);
\draw (0.6,-3) -- (0.6,-4);
\draw (1,-3) -- (1,-2.8);
\draw (2,-3) -- (2,-2.8);
\draw (2.4,-3) -- (2.4,-4);
\draw (0.6,-4) -- (2.4,-4);
\draw [stealth-](2.4,-3.4) -- (3,-3.4) node[right] {Stabilising rig};
\draw[fill=gold!90] (0.6,-4) -- (0.2,-4) -- (0.2,-3.8) -- (0.6,-3.8);
\draw[fill=gold!90] (2.4,-4) -- (2.8,-4) -- (2.8,-3.8) -- (2.4,-3.8);
\draw[fill=gold!90] (1.3,-4) -- (1.3,-3.8) -- (1.7,-3.8) -- (1.7,-4);

\draw [-,line width=0.3mm](2.8,-3.86) -- (3.6,-3.86) node[right] {BaSyTec channel};
\draw [-,red,line width=0.3mm] (2.81,-3.9) -- (3.6,-3.9);
\draw [-,blue,line width=0.3mm] (2.81,-3.94) -- (3.6,-3.94);

\draw[fill=gray!40] (1.1,-7) rectangle (1.9,-4);
\node at (1.5,-6.7) {$-$};
\node at (1.5,-4.3) {$+$};

\draw [pattern=north east lines,pattern color=black] (0.7,-7) rectangle (1.1,-4);
\draw [pattern=north east lines,pattern color=black] (2.3,-7) rectangle (1.9,-4);

\draw (2.3,-4.4) rectangle (2.4,-6.7);
\draw (0.6,-4.4) rectangle (0.7,-6.7);

\draw [fill=gray!40] (2.4,-4.4) rectangle (2.6,-6.7);
\draw [fill=gray!40] (0.4,-4.4) rectangle (0.6,-6.7);

\draw [fill=gray!80] (0,-4.4) rectangle (0.4,-6.7);
\draw [fill=gray!80] (2.6,-4.4) rectangle (3,-6.7);
\filldraw [fill=blue!30] (0.2,-4.9) circle (0.13);
\filldraw [fill=blue!30] (0.2,-6.2) circle (0.13);
\filldraw [fill=blue!30] (2.8,-4.9) circle (0.13);
\filldraw [fill=blue!30] (2.8,-6.2) circle (0.13);

\draw [fill=gray!70] (0.2,-7) rectangle (2.8,-7.3);

\draw (-0.4,-7.3) rectangle (3.4,-7.3);
\draw [pattern=north east lines,pattern color=black,draw=none] (-0.4,-7.3) rectangle (3.4,-7.6);

\draw[stealth-] (1.9,-5.2) -- (3.6,-5.2) node[right] {Cell};
\draw[stealth-] (2.3,-5.8) -- (3.6,-5.8) node[right] {Aluminium block};
\draw[stealth-] (2.6,-6.5) -- (3.6,-6.5) node[right] {Peltier element};
\draw[-stealth] (-0.5,-4.9) node[left] {Coolant channel} -- (0.1,-4.9) ;
\draw[-stealth] (-0.5,-5.5) node[left] {Coolant block} -- (0,-5.5) ;
\draw[-stealth] (-0.5,-6.5) node[left] {Thermal interface} -- (0.6,-6.5) ;


}
\end{tikzpicture}

%% file: Sections/new_Results_and_Discussion.tex
\section*{Results and Discussion}

In this section, several variants of the standard Th\'evenin ECM, parameter identification, and SOC estimation processes are introduced, as summarised below:

\begin{itemize}
    \item \textbf{ECM:} standard Th\'evenin ECM, with ordinary least-squares parameterisation;
    \item \textbf{ECM-opt:} ECM with nonlinear optimisation for parameter identification;
    \item \textbf{ECMh:} ESC (Th\'evenin ECM with a hysteresis component) and ordinary least-squares parameterisation;
    \item \textbf{ECMh-opt:} ECMh with nonlinear parameter optimisation;
    \item \textbf{Constant-EKF:} extended Kalman filter SOC estimator, with \(\Sigma_v\) and \(\Sigma_w\) constant across depth of discharge;
    \item \textbf{Adaptive-EKF:} EKF with \(\Sigma_v\) and \(\Sigma_w\) varying according to estimated SOC, and current draw \(i(t)\).
\end{itemize}

\begin{figure}

    \begin{subfigure}{0.95\textwidth}
    \centering
    \includegraphics[width=\textwidth]{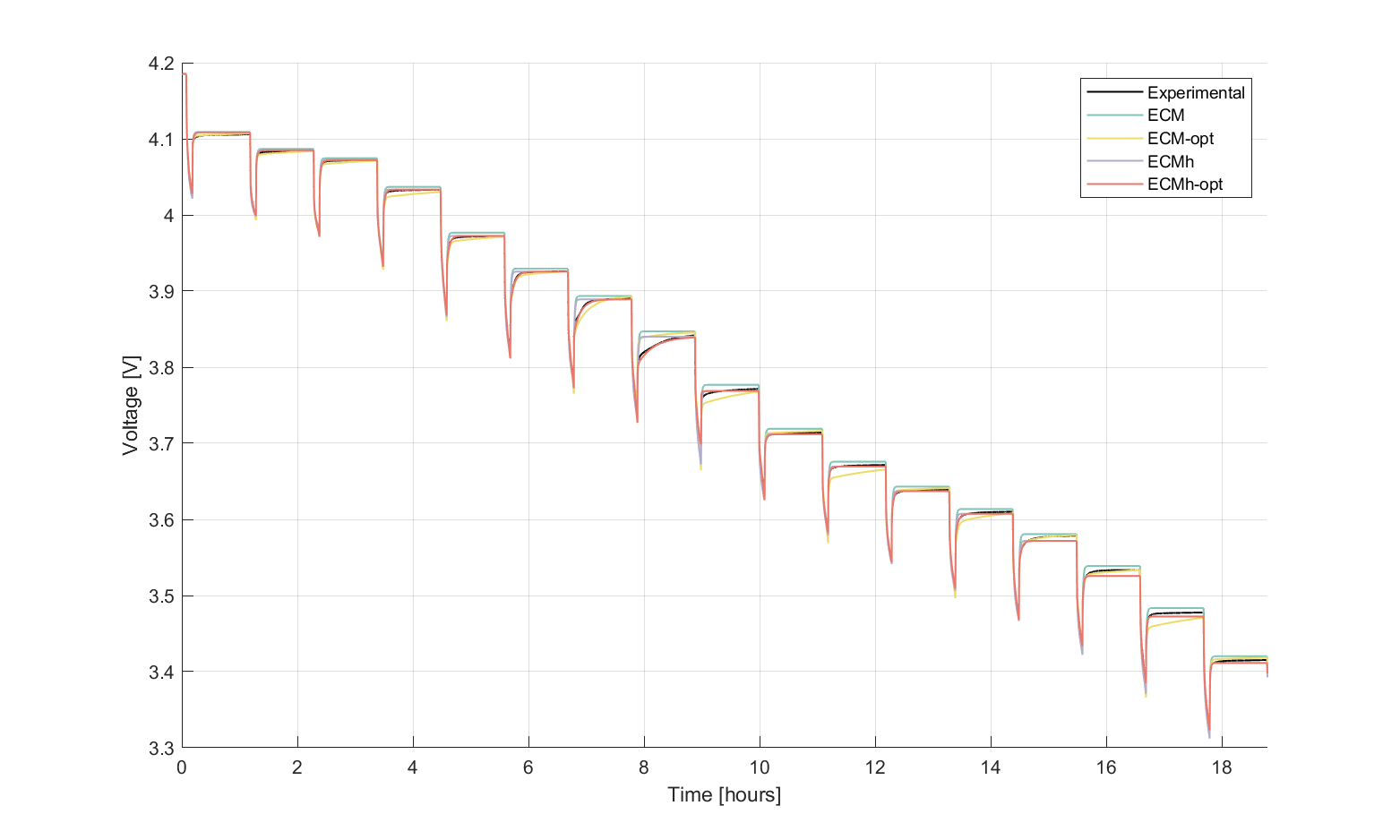}
    \caption{Fitted voltage}
    \end{subfigure}

    \begin{subfigure}{0.95\textwidth}
    \includegraphics[width=\textwidth]{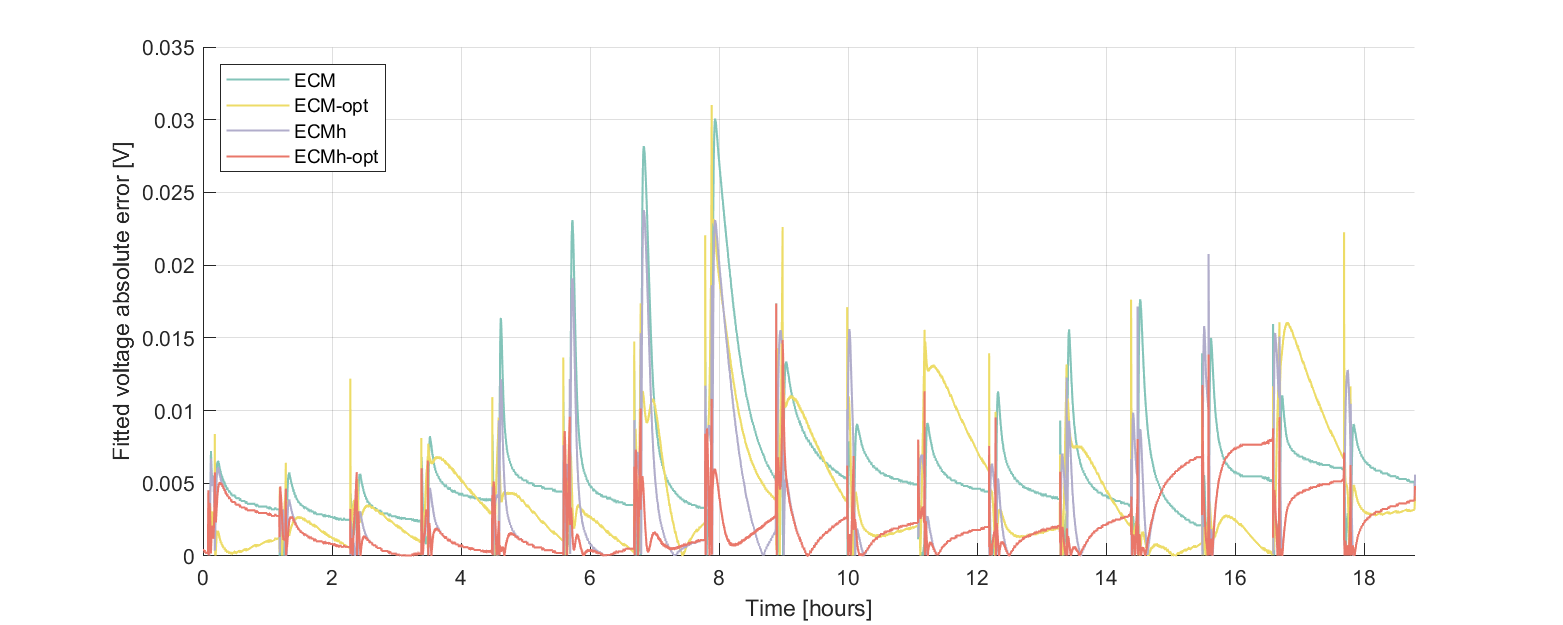}
    \caption{Fitted voltage absolute error}
    \end{subfigure}

    \caption{Experimentally measured voltage from the GITT experiment on Cell A, compared to the fitted voltage for the ECM, with and without parameter identification optimisation and with and without hysteresis}
    \label{fig:GITT fitting example}
\end{figure}

Figure~\ref{fig:GITT fitting example} illustrates the results of the fitting procedures outlined in the methodology section.
Model fits are shown on GITT data from cell A, which is used to parameterise a heirarchy of increasingly accurate models.
Firstly, a Th\'evenin model is parameterised from the data, labelled `ECM', using ordinary least-squares with \(\tau_1=2\) s and \(\tau_2=50\) s.
Next, the ECM parameters are refined using the numerical optimisation routine discussed in the methodology section.
Circuit component paramters, including timescales \(\tau_1\) and \(\tau_2\), are fitted to the observed data to give a better model fit, labelled `ECM-opt'.
Separate charge and discharge OCV curves are computed from GITT-for-OCV data, which are used in conjunction with ordinary least-squares parameterisation and the ESC model, giving a hysteretic ECM labelled `ECMh'.
Finally, GITT and GITT-with-charge-pulse data are used to optimise the ECMh parameter values, labelled `ECMh-opt'.
Each step results in a model whose structure or parameters capture more battery dynamics than the previous model, resulting in increasing accuracy.
For Cell A, the RMS voltage-fitting error is 7.0 mV for ECM, 6.0 mV for ECM-opt, 4.9 mV for ECMh and 3.0 mV for ECMh-opt. 

\begin{figure}

    \begin{subfigure}{0.95\textwidth}
    \centering
    \includegraphics[width=\textwidth]{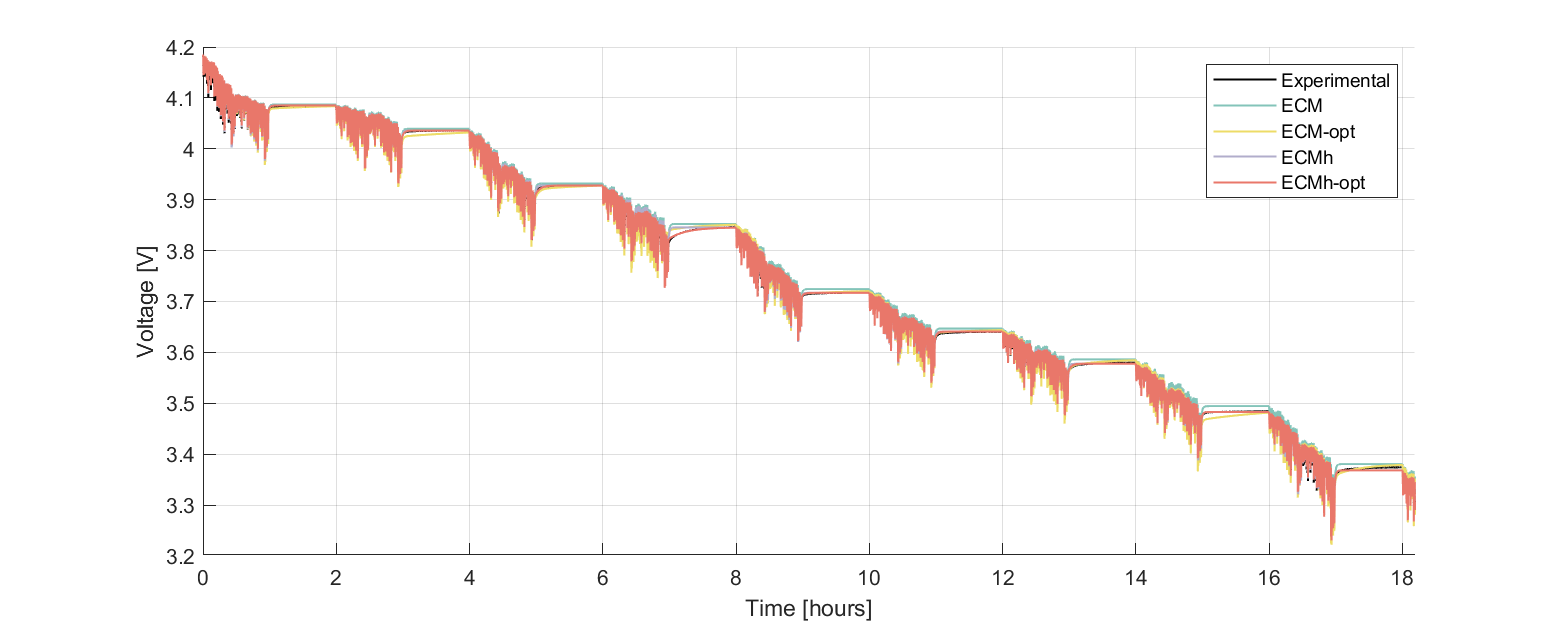}
    \caption{Predicted voltage.}
    \end{subfigure}

    \begin{subfigure}{0.95\textwidth}
    \includegraphics[width=\textwidth]{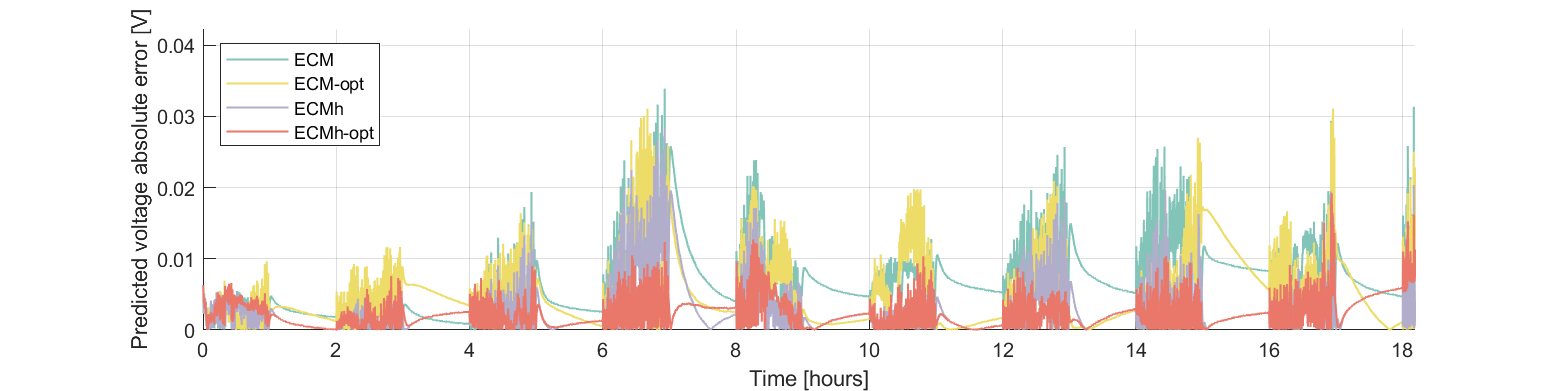}
    \caption{Predicted voltage absolute error.}
    \label{sfig:WLTP Error Cell A}
    \end{subfigure}

    \caption{Experimentally measured voltage from the WLTP experiment on Cell A, compared to the fitted voltage for the ECM, with and without parameter identification optimisation and with and without hysteresis.}
    \label{fig:WLTP Cell A}
\end{figure}

\begin{figure}
    \centering
    \includegraphics[width=0.95\textwidth]{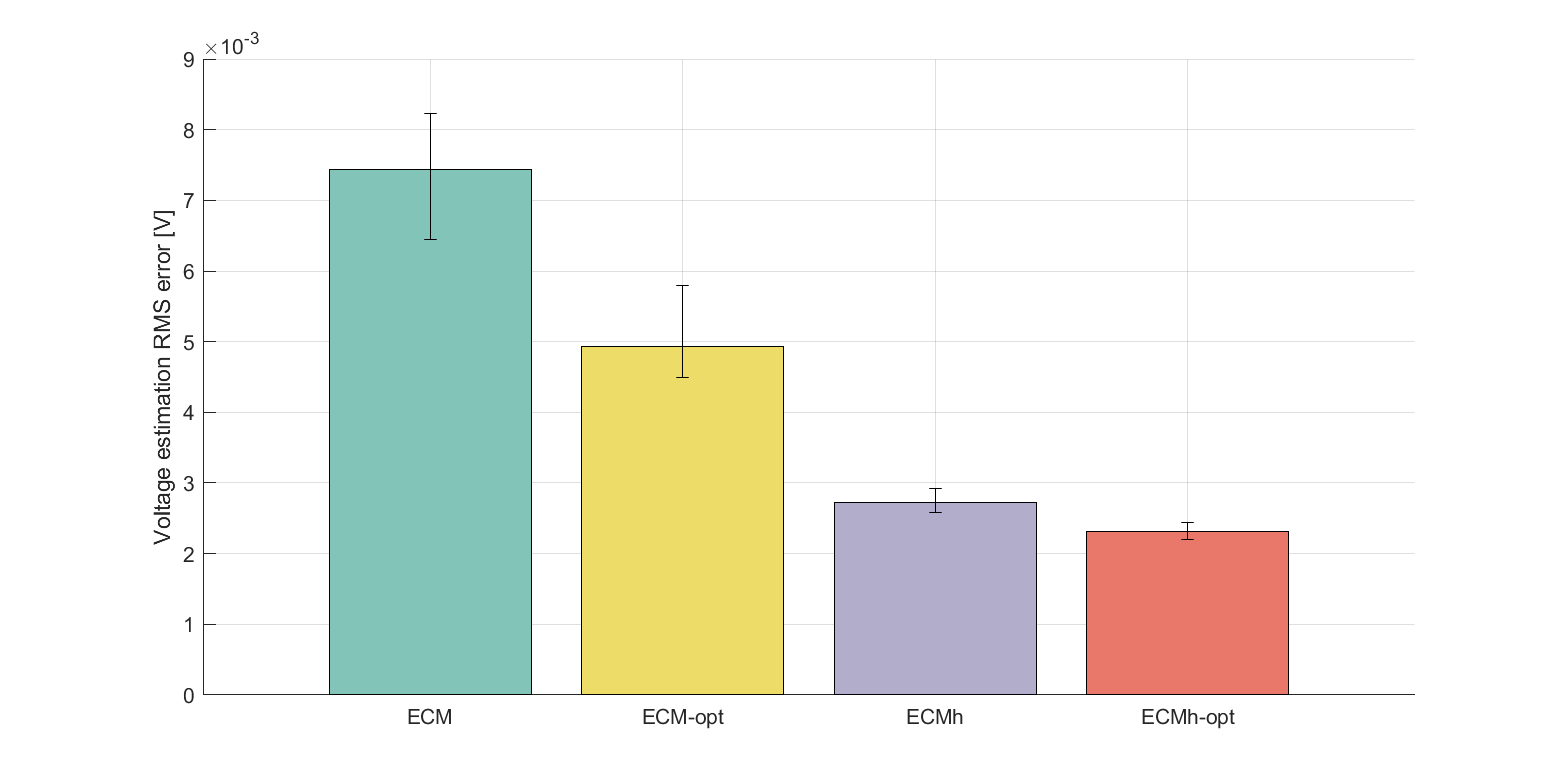}
    \caption{The mean RMS voltage prediction error for all models of each cell tested, evaluated against the WLTP drive cycle. Error bars report the maximum and minimum RMS voltage prediction error for each of the models}
    \label{fig:WLTP RMSE voltage}
\end{figure}

Figure \ref{fig:GITT fitting example} provides evidence that the optimised parameter identification process and the hysteresis component result in better fits to the parameterisation data. 
Nevertheless, comprehensive validation requires the model performances to be evaluated against an independent dataset. 
Validation results are shown in Figure \ref{fig:WLTP Cell A}, where each model is evaluated on the developed WLTP drive cycle. Time series of voltage predictions and absolute errors are plotted.
In addition, the RMS voltage errors from the WLTP validation process are displayed in Figure \ref{fig:WLTP RMSE voltage}, which shows the average RMS errors over all cells, with maximum and minimum cell RMS errors displayed as error bars.

Benefits are clearly obtained from the optimisation-based parameterisation, and the inclusion of a hysteresis component.
Significant reductions in absolute error are observed in both the dynamic periods, where load is on the cell, and during periods of rest.
Across all cells, the same common trend is present --- optimising parameter identification processes or including a hysteresis component within the ECM will enhance the performance of the model.
Furthermore, error bars are seen to shrink as model accuracy improves.
This indicates that while the performance of simple ECMs can fluctuate substantially from cell to cell, hysteresis elements and better parameterisation creates more robust models.

It is noted that a small steady state error remains; when using ECMh-opt, the predicted return to OCV carries more error at some SOCs than ECM-opt.
This was consistent across all cells, suggesting that small errors exist in the parameterised OCV curves.
Such errors can arise from the GITT for OCV procedure, where rest times are too short for the cell to reach equilibrium.
This suggests that yet higher accuracy may be achievable if rest-times are further increased during parameterisation experiments.
OCV LUTs could also be included in the optimisation procedure, so that the numerical optimiser can correct errors in experimental OCV measurements.
Results also suggest that the parameter optimisation process in ECM-opt can help to eliminate rest-OCV error. 
Nevertheless, parameters that are optimised to eliminate rest-OCV error are less accurate at describing kinetic periods, evidenced from the data shown in Figure \ref{sfig:WLTP Error Cell A}.
This results in the increased magnitude of the ECM-opt error bar in Figure \ref{fig:WLTP RMSE voltage}.
To overcome this, separate resting and dynamic parameter tuning could be investigated, using decoupled optimisations for kinetic and rest periods.
While this is an interesting avenue for further cell model improvements, we consider it to be beyond the scope of the present study, since it would bring a third parameter, current \(i(t)\) into the LUT.

\subsection*{SOC estimation and Adaptive Extended Kalman Filter}

Extended Kalman filter (EKF) SOC estimation is validated on the developed WLTP drive cycle data.
The following discussion is set against the performance criteria most relevant to the BMS industry --- SOC estimation. 
This allows focus to be set directly on the performance of the EKF tuning methods applied to the various models.
For completeness, the exact models discussed in the previous section are used to establish a baseline estimation accuracy, without any EKF inclusion.
As the WLTP drive cycle is power-based, this is achieved by using a given model to compute the necessary current draw satisfying the power requirements at any given time-step, and Coulomb-counting on the modelled current-draw to estimate SOC. 
The method is highly computationally efficient but has no mechanism to access feedback from true measurements made on the cell, meaning it is prone to SOC-drift over a period of time. 
The EKF tuning methods are tested on the drive cycle using a similar approach, whereby the chosen model is used to calculate the necessary current-draw at each time step in the drive cycle. 
However, modelled current-draw is then input to an EKF (with specified tuning) for SOC estimation, rather than Coulomb counting as used for baseline estimation accuracy.

\begin{figure}

    \begin{subfigure}{0.95\textwidth}
    \centering
    \includegraphics[width=\textwidth]{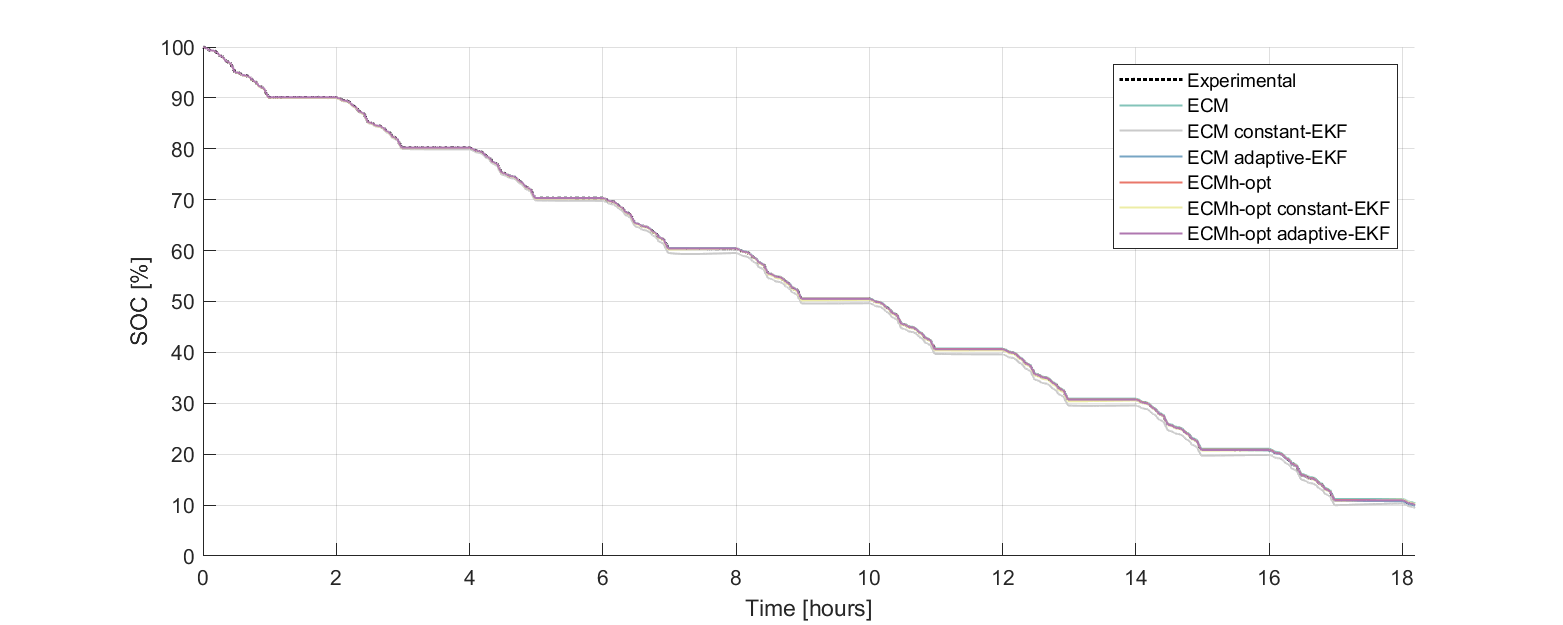}
    \caption{Predicted SOC - baseline Coulomb counting, constant Kalman filter and adaptive Kalman filter}
    \label{sfig:SOC Error Cell A}
    \end{subfigure}

    \begin{subfigure}{0.95\textwidth}
    \includegraphics[width=\textwidth]{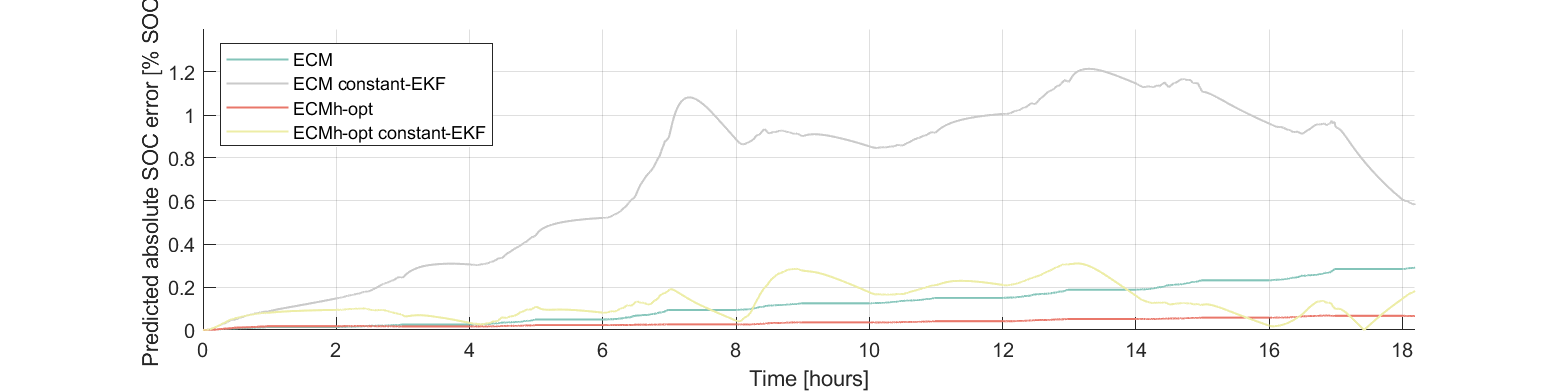}
    \caption{Predicted SOC absolute error  - baseline Coulomb counting and constant Kalman filter}
    \label{sfig:SOC RS Error Cell A - constant}
    \end{subfigure}

    \begin{subfigure}{0.95\textwidth}
    \includegraphics[width=\textwidth]{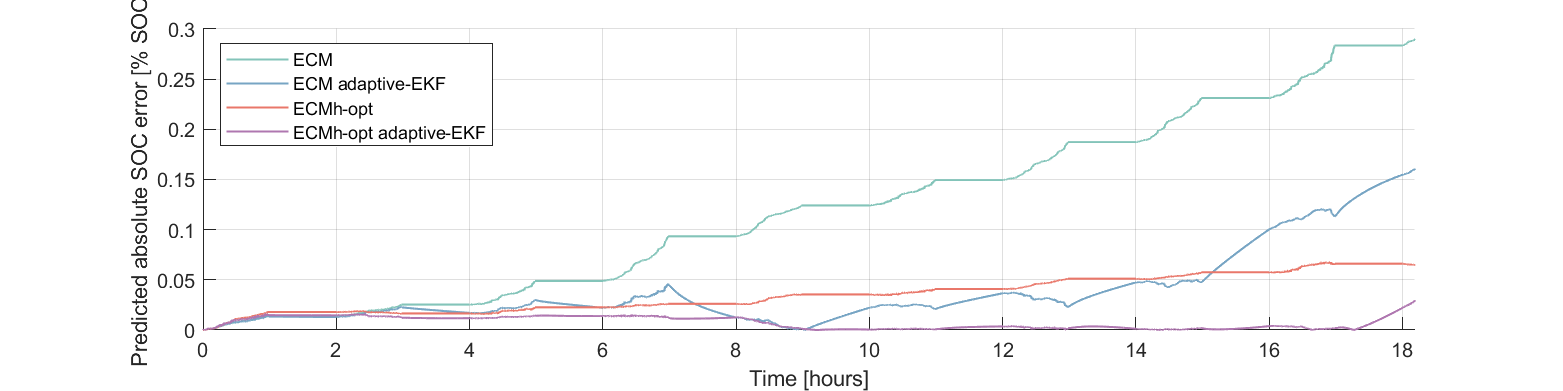}
    \caption{Predicted SOC absolute error  - baseline Coulomb counting and adaptive Kalman filter}
    \label{sfig:SOC RS Error Cell A - adaptive}
    \end{subfigure}

    \caption{Experimentally measured SOC from the WLTP experiment on Cell A, compared to the predicted SOC from ECM and ECMh-opt, through Coulomb counting, constant and adaptive Kalman filtering} 
    \label{fig:SOC Error Cell A}
\end{figure}

\begin{figure}
    \centering
    \includegraphics[width=0.95\textwidth]{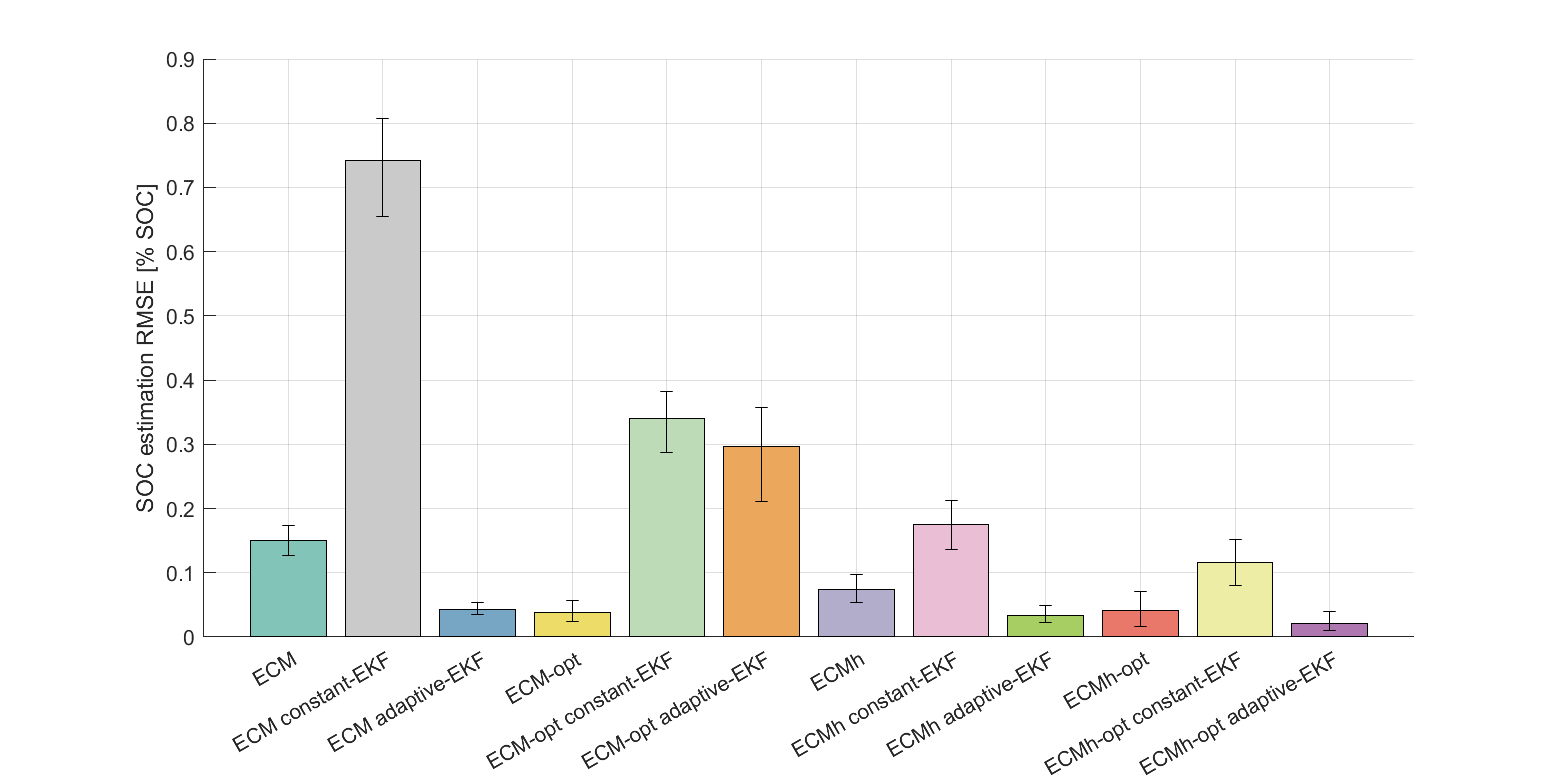}
    \caption{The mean SOC prediction error for all models of each cell tested, evaluated against the WLTP drive cycle. Error bars report the maximum and minimum RMS SOC prediction error for each of the models}
    \label{fig:WLTP RMSE}
\end{figure}

Each model is tested with both constant and adaptive EKF tuning, and compared against true cell SOC calculated via Coulomb counting.
Results are depicted in Figure~\ref{fig:SOC Error Cell A}, showing time-series results for cell A with the ECM and ECMh-opt models, and in Figure~\ref{fig:WLTP RMSE}, depicting RMS error in SOC estimation for every combination of model and EKF tuning.
Figure~\ref{fig:WLTP RMSE} shows the mean RMS error in SOC estimation computed across all cells, with error bars indicating the cells with the smallest and largest estimation errors.

Observing Figure \ref{fig:WLTP RMSE}, the Coulomb counting method is shown to outperform the constant-EKF method for all introduced model architecture and parameter tuning methods. 
This highlights a clear limitation in the constant parameter EKF method - the additional computation required by this method does not offer a reduction in SOC estimation performance. 
Focusing on the specific case of Cell A, the poor performance of constant-EKF is clear in Figure \ref{sfig:SOC RS Error Cell A - constant}.
However, the Coulomb counting method is also seen to be limited in Figure \ref{sfig:SOC RS Error Cell A - adaptive}.
The magnitude of SOC estimation error can be seen to increase as the drivecycle progresses (for both ECM and ECMh-opt), to the extent that an SOC estimation error of 0.29 \% exists by the end of the 18-hour cycle.
This is likely caused by the lack of feedback mechanism when Coulomb counting through the model and it is reasonable to expect that this error will continue to grow over longer drive cycles and repeated charge-discharge cycles.
This problem could be alleviated by integrating a feedback mechanism into the model, for example, by cross-referencing against a SOC-OCV LUT table at regular intervals. However, this would increase the complexity of the SOC estimation method and reduce the attraction of the lightweight modelling method. 

The adaptive-EKF tuning is shown to perform very effectively for the ECM, ECMh and ECMh-opt models, reducing SOC estimation error in all cases (Figure \ref{fig:WLTP RMSE} and, crucially, mitigating the issue of drift that limits the Coulomb counting method, evidenced in Figure \ref{sfig:SOC RS Error Cell A - adaptive}). 
Firstly, this emphasises the importance of high quality, adaptive-EKF tuning for effective SOC estimation in both standard ECMs and ECMs including hysteresis components. 
Secondly, this highlights that for the NMC811 cells used in this study, the development of effective adaptive-EKF tuning methods should be considered a higher priority than the inclusion of hysteresis components and development of hysteresis parameterisation methods. 
With the most basic ECM, the adaptive-EKF tuning method is able to significantly reduce SOC estimation error, below what is achieved with the inclusion of hysteresis terms and parameter optimisation methods. 
For other cell chemistries, particularly LFP cells which are known to have significant hysterical components within their OCV, this statement must be stress tested before the development of SOC estimation methods for a BMS.

The simplest model tested here is the ECM without hysteresis, parameterised using ordinary least-squares.
Counterintuitively, the combination of this ECM with an adaptive EKF outperforms the optimised ECM for SOC estimation.
As time constants \(\tau_j\) are allowed to change during the optimisation procedure, there is more variation between the parameters of each ECM-opt model than there is in the parameters of the ECM model.
Higher parameter variances causes the EKF to assume greater process noise. As a result, the EKF has greater uncertainty in the ECM-opt SOC estimate and, despite its better parameter set, places less weight on the model results. Thus, greater fluctuation in SOC estimation error is expected, due to the EKF underweighting model predictions whilst filtering, and relying more on noisy measurements. This further suggests that accurate characterisation of parameter variances is critical to adaptive EKF accuracy.
Given that sample variance is a random variable, a more accurate estimate of parameter variances would be obtained if more than three cells were tested.
Ideally, a larger number of cells would be parameterised, resulting in a more meaningful estimate of the parameter variances and process noise covariance matrix. 
Whilst the performance of the adaptive-EKF SOC estimation methods is very strong in this study, a greater number of cells would certainly make the process more robust, with reduced chance of parameter error caused by singular erroneous results.

%% file: Sections/Conclusions.tex
\section*{Conclusions}
\label{sec:Conclusions}

Parameterisation of lithium-ion cell equivalent circuit models is well-established.
Here, these methods are generalised to parameterise models with hysteresis, with the introduction of short charge steps to decouple hysteresis from other electrochemical phenomena observed during the completion of a kinetic parameterisation procedure (GITT).
Results demonstrate that a well-parameterised hysteresis model reduces RMS voltage prediction error by approximately 50 \% on relevant test data, when compared to ordinary least-squares parameterisation of models without hysteresis, as commonly employed in the literature.

Accurate cell models are desirable for control purposes in BMSs, with SOC estimation being a key motivation.
Adaptive extended Kalman filtering has been proposed for estimating SOC in models with hysteresis~\cite{Rzepka}, however the method has not previously been validated experimentally.
This important validation is provided here, which is used to demonstrate that a well-parameterised hysteresis model provides very high SOC estimation accuracy when paired with an adaptive extended Kalman filter.

Results show that better cell models will typically yield more accurate SOC estimates.
Furthermore, improvements in the Kalman filter, through inclusion of adaptive covariances, always leads to improvements in SOC estimation.
Improvements to the Kalman filter are seen here to produce more substantial increases in SOC estimation accuracy, compared to improvements that can be gained by enhancing the underlying model with hysteresis components and parameter optimisation --- adaptive covariance matrices reduce SOC estimation error by up to 85 \%.
It is important that this statement is tested on other cell chemistries, particularly those that are known to have a considerable hysterical component within their OCV, such as LFP cells.

Future work should focus on refinements to extended Kalman filter estimation.
A key step towards this is accurate estimation of parameter variances, when generating the process noise covariance matrix of an adaptive extended Kalman filter.
A further avenue to explore is the inclusion of an explicit temperature dependency in the cell models and associated Kalman filters.
Precise thermal control is unachievable in EV applications; future work should therefore extend adaptive Kalman filters to include the effects of cell temperature and heat generation in the process noise covariance matrix.
As a typical EV battery pack contains many individual cells, efforts should also be made to scale adaptive-EKF methods up to pack-level SOC estimation.

%% file: Sections/Acknowledgements.tex
\section*{Acknowledgements}

The authors disclose support for this work from the Faraday Institution (grant number FIRG059).

%% file: Sections/Author_Contributions.tex
\section*{Author contributions}
\label{sec:Author contributions}

All authors contributed equally towards the development of the research, the research strategy and the design of experiments. J.K. led the development of all equivalent circuit models and all parameter identification methodologies set out in this study. J.K led the initial drafting of the manuscript and contributed heavily towards editing of the manuscript, prior to submission. M.B. contributed towards the development of the research strategy and led the experimental work set out in this study. M.B. contributed towards the development of the the equivalent circuit models and the parameter identification methodologies. M.B. contributed to the initial drafting of the manuscript and led the editing/ additional drafting of the manuscript. A.H. contributed towards experimental work and the parameter identification methodologies set out in this study. A.H. contributed towards the editing of the manuscript. 

%% file: Sections/Data_Availability.tex
\section*{Data Availability}

The data that support the findings of this study are available from the corresponding author upon reasonable request.

%% file: Sections/Code_Availability.tex
\section*{Code Availability}

The code that support the findings of this study are available from the corresponding author upon reasonable request.

%% file: Sections/Competing_Interests.tex
\section*{Competing interests}
\label{sec:Competing interests}

The authors declare no competing interests.